# Structural evolution of MnTi$_{0.8}$Ru$_{0.2}$O$_3$

R. K. Maurya*

*School of Basic Sciences, Indian Institute of Technology Mandi, Kamand, Himachal Pradesh-175005, India.*

**Abstract:** Here We present the structural studies on MnTi$_{1-x}$Ru$_x$O$_3$(x=0, 0.2) compounds. The role of Ti ions in the magnetism of MnTiO$_3$ was unclear so here this issue has been tried to address in this manuscript. The magnetic susceptibility data shows that the 3-dimensional magnetic character has been improved in the doped MnTiO$_3$. The x=0 compound goes paramagnetic to antiferromagnetic phase at a temperature ~64K followed by a broad peak at a temperature ~100K . But in x=0.2 compound the antiferromagnetic transition temperature has been shifted towards the lower temperature to ~ 47.5K with a broad peak at temperature ~79K. On the doping of Ru at Ti site a sharp anomaly is observed ~47.5K in the case of x=0.2 compound. This sharp anomaly attributes to the improved 3D character of magnetism in this compound which is weak in x=0 compound.

**Introduction:**

$MnTiO_3$ stabilises in the ilmenite structure with the centrosymmetric space group $R\bar{3}$ in which $Mn^{2+}$ and $Ti^{4+}$ ionic layers are arranged alternatively along the hexagonal c-axis. In the 3D crystal structure of $MnTiO_3$, the same type of octahedra (i.e. $MnO_6$ or $TiO_6$) are connected to each other by edge sharing and the different types of octahedra ( i.e. $MnO_6$ and $TiO_6$) are connected to each other by face sharing. The cations form a distorted honeycomb pattern in the c-plane, arranged alternatively along the hexagonal c-axis. The distortion in the cationic honeycomb pattern comes because of the electrostatic repulsion between the cations along the c-axis. This compound shows interesting properties. The thin film of $MnTiO_3$ shows ferrotoroidicity[1]. The linear magnetoelectric effect has been observed in single crystals of $MnTiO_3$[2]. $MnTiO_3$ shows quasi- two dimensional antiferromagnetic behaviour because of the accidental cancellation of different inter-layer exchange interactions[3,4]. In this system, the magnetism arises because of the $Mn^{2+}$ ions having partially filled d-orbitals($3d^5$). The magnetic properties of some powder ilmenites viz. $MnTiO_3$, $FeTiO_3$, $CoTiO_3$ and $NiTiO_3$ have been studied by J.J Stickler *et al*[5]. In this study they exhibited the behaviour of magnetic susceptibility measurement with temperature. $MnTiO_3$ shows a broad peak around 100K with $T_N$ around 64K. But other ilmenites $FeTiO_3$, $CoTiO_3$ and $NiTiO_3$ exhibit a sharp peak at $T_N$ as compared to $MnTiO_3$. Neutron diffraction studies have also been done on powdered $MnTiO_3$ and $NiTiO_3$ by G. Shirane *et al*[6].The neutron diffraction studies revealed that the spin arrangements in $MnTiO_3$ are different from $NiTiO_3$. In $MnTiO_3$ the spins are arranged antiferromagnetically along the hexagonal c-axis as well as in basal plane while the spins are arranged ferromagnetically in the c-plane in case of $NiTiO_3$. In $FeTiO_3$, the spin arrangement is similar to $NiTiO_3$.These different types of arrangement of spins leads to the exotic spin glass behaviour in the certain doped compounds such as $Mn_{1-x}Ni_xTiO_3$ and $Fe_{1-x}Mn_xTiO_3$[7,8]. The magnetic and electrical properties of $MnTi_{1-x}Nb_xO_3$ have been studied by Ramakrishnan *et al.* for certain value of doping amount x[9]. In this study they showed the improvement in the three dimensional magnetic character with the doping of Nb at Ti site. The magnetic properties of some ilmenites have been studied by Goodenough *et al*[10]. Here they proposed five superexchange interactions to explain the 2D and 3D antiferromagnetism in $MnTiO_3$. Among the five exchange interactions 2 are intra-layer and 3 are inter-layer. The intra-layer interactions are responsible for the 2D magnetism and inter-layer for 3D magnetism in this system. $MnTiO_3$ shows a spin-flop transition at an applied magnetic field of ~ 6.5T in addition to spin canting[11] . Mufti et al showed that in $MnTiO_3$ there is a weak spin-lattice coupling[2]. They suggested this fact on the basis of non-appearance of dielectric anomaly in the absence of applied magnetic field . The non-appearance of dielectric anomaly due to spin lattice coupling indicates that the nature of the spin structure is such that no break in spatial symmetry observed. But in our recent study we found that in $MnTiO_3$ the spin lattice coupling is not weak in the absence of external magnetic field[12]. Hence the nonappearance of dielectric anomaly in $MnTiO_3$ does not mean that the spin lattice coupling is weak. In the recent study done by H. J. Silverstein *et al.* it has been proposed that the local spin lattice coupling is not weak and it is much more important for the magnetoelectric properties of this compound[13]. In our previous study[12], the Mn-O, Mn-Mn and Ti-O bonds are showing significant changes across the region of intra- and inter-layer antiferromagnetic interactions. The changes in Mn-O and Mn-Mn bonds across the onset temperature for 2D and 3D AFM is understandable but the changes in the Ti-O bonds at these temperature regions is not understandable . Hence the role of $Ti^{4+}$ ions is unclear in the magnetism and the magnetically induced ferroelectricity in $MnTiO_3$. This issue motivated us to investigate this compound. To observe the change in the magnetism of this compound, the Ru ions have been doped at Ti site. The doping of Ru ions in a very small amount at Ti site changes the magnetic properties substantially. Our magnetic susceptibility results show that this doping improved the sharpness in magnetic transition peak with the shift toward the lower temperature and a shift in the broad peak toward the lower temperature. The $T_N$ and $T_{2D}$ have been changed severely because of this doping. The presence of sharp anomaly at $T_N$ suggests the improvement in the three dimensional antiferromagnetic character. In x= 0 compound, there is no significant anomaly at the magnetic phase transition temperature. The cause of insignificant anomaly at $T_N$ is the accidental cancellation of different inter-layer exchange interactions[10]. Our x-ray diffraction results show that he behaviour of lattice parameters has been affected by the doping of Ru ions. In x=0 compound, the lattice parameters are linear up to 200K but in the case of x=0.2 compound, the linear behaviour is up to

225K which indicates that the onset temperature for two dimensional antiferromagnetic interactions has been shifted towards the higher temperature by a temperature of 25K.

**Experiment:** The polycrystalline samples of $MnTiO_3$ and $MnTi_{0.8}Ru_{0.2}O_3$ were prepared using the conventional solid state route. The stoichiometric amounts of $MnCO_3$, $TiO_2$ and $RuO_2$ were mixed and ground using mortar and pestle. The mixtures were heated at 1200°C for 24 hours in air. The samples were characterized using the powder x-ray diffractometer, DC magnetization techniques. The temperature dependent powder XRD measurements were carried out using the Smart Lab 9 kW rotating anode x-ray diffractometer with Cu $K_\alpha$ radiation. The XRD data were collected from 300K to a temperature down to 26K with the help of closed cycle cryogenic helium compressor unit. The measurements were carried out for 16 different temperatures (300K to 26K) for x=0.2 compound. These XRD patterns were analysed with Rietveld Refinement method using the FullProf software[14]. The DC magnetization measurement were carried out using MPMS 3 Quantum Design magnetometer in the temperature range 300K to 2K at an external magnetic field of 0.1Tesla.

**Results and discussion:**

The x-ray diffraction patterns of x=0.2 compound down to 26K show that it stabilises in the centro-symmetric space group $R\bar{3}$ with the lattice parameters close to x=0 compound(for the sake of comparison the published results of $MnTiO_3$ have also been included throughout the paper). Therefore the crystal structure of x=0 compound was used as the starting model for x=0.2 compound. Figure1, presents the Rietveld refinement of x=0.2 compound collected at 300K and 26K. The room temperature and 26K data were indexed using the $R\bar{3}$ space group with the lattice parameters a=5.1330(9) Å, c=14.2748(3)Å and a=5.1249(9)Å, c=14.2638(3)Å, respectively. The goodness of fit parameter S, for both the temperatures is ~1.6. The results indicate the complete solid solution for the doping content x= 0 and 0.2 . The values of lattice parameters are in good agreement with previous reports[2,12]. Both the lattice parameters a and c show a decrement with the Ru doping. The sample does not show any structural phase transition when it was cooled down to 26K. The shift in the peak position is observed towards the higher 2θ(deg.) value due to lattice contraction on decreasing the temperature.

To extract the different structural parameters across the magnetic phase transition , the XRD patterns at all the temperatures were analysed using FullProf . Figure 1 shows, typical Rietveld refinement of the XRD patterns for x=0.2 compound. The goodness of fit parameter S, obtained for all the temperatures is close to 1.6. The goodness of fit parameter S, for x=0 compound is close 1.5[9]. The temperature evolution of lattice parameters a and c for x=0 and x=0.2 compounds are shown in Fig2(a) and 2(b), respectively. The value of lattice parameters a and c decreasing with the doping amount. The behaviour of lattice parameters a and c with the temperature has been shown in figure 2. On reducing the temperature the reduction in lattice parameters is not linear for both the compounds in the whole temperature range. By the careful observation for x=0.2 system, it is observed that both the lattice parameters a and c do not show a linear behaviour for the whole temperature range. The lattice parameters are linear in the temperature region from 300K to 225K. After 225K these parameters show deviation from the linearity. This deviation is due to the competing intra-layer antiferromagnetic interactions set in~225K. The decrement in lattice parameters a and c in the temperature range 300K to 225K is 0.05% and 0.07%, respectively. In the temperature range 225K to 125K both the lattice parameters a and c are decreasing by 0.049% and 0.052%, respectively. In the temperature range 125K to 26K the lattice parameter a is showing a decrement of 0.048% while the lattice parameter c shows an increment by 0.044%. So, here the lattice parameter a is showing a very less deviation from the linear thermal behaviour up to the lowest temperature 26K. But for x=0 compound the situation is different[12].

Since the ionic radius of $Ti^{4+}$ ion (0.605Å) is smaller than the $Ru^{4+}$ ion (0.620Å) hence on increasing the doping content, the lattice parameters should increase but here these are decreasing with doping. This is a puzzling result. Why is this happening has to understand. For x=0 compound, the decrement in lattice parameters a and c is linear up to 200K but for x=0.2 compound the decrement is linear up to 225K. For x=0.2 compound, the lattice parameter c becomes minimum~125K after that it is increasing continuously down to the lowest temperature. For x=0.2 compound, below 125K the increment in lattice parameter c is 0.044% while for parent compound below 95K the increment is 0.014%. Hence the lattice parameter c have a sharp increment by 0.03% more than the x=0 compound which is numerically much significant.

Fig.3 illustrates the behaviour of lattice parameters with the temperature for x=0 and x=0.2 compounds. For x=0.2 compound, in the temperature range 300K to 225K the lattice parameters show a linear behaviour. In the temperature region 225K to125K, the lattice parameters show a deviation from linearity.

Figure 4 shows the behaviour of c/a ratio with the temperature. The c/a ratio is increasing with increase of doping content which indicates that the difference between the lattice parameters is decreasing. But there is no structural distortion in these compounds. For x=0, the c/a ratio decreases on reducing the temperature and it gets a minimum value~140K. Further on reducing the temperature the c/a ratio increasing with the decrement in temperature which indicates the separation between the two intervening oxygens decreases on the cost of increment in Mn-O(s) bonds. The reduction in the separation between these two oxygen leads to the enhancement in the super superexchange interaction between $Mn^{2+}$ spins via Mn-O-O-Mn path.

On the other hand, in case of x=0.2 compound, the c/a ratio decreasing on reducing the temperature and it becomes minimum~225K. The c/a ratio has constant value in the temperature region of 225K to 125K after that it is increasing rapidly down to the lowest temperature. This constant temperature region for c/a ratio is because of almost same variation in both the lattice parameters. The reduction in lattice parameters, a and c in the temperature range 225K to 125K is 0.44% and 0.5% which is almost same.

To understand the temperature dependent behaviour of lattice parameters we have subtracted the thermal contribution of the lattice parameters from the raw data. Figure 5 shows, the linear thermal contribution subtracted lattice parameters a and c. Now these parameters have only magnetic contribution. For x=0, both the lattice parameters a and c are increasing below the 200K while in case of x=0.2 compound parameters start increasing below 225K. In x=0.2 compound the linear behaviour of lattice parameters a and c has shifted towards the higher temperature by a temperature of 25K as compared to x=0 compound i.e. the onset temperature for the 2 dimensional magnetism is shifted by 25K towards the higher temperature.

Figure 6, illustrates the zero field cooled (ZFC) and field cooled (FC) magnetic susceptibilities of x=0 and x=0.2 compounds at an external magnetic field of 0.1 tesla. The x=0 compound shows a paramagnetic to antiferromagnetic transition $T_N$ ~ 64K followed by a broad peak at $T_{2D}$ ~100K. Here the $T_{2D}$ represents the paramagnetic to two dimensional antiferromagnetic transition temperature and $T_N$ for paramagnetic to three dimensional AFM transition temperature. This broad peak at a temperature~ 100K attributes to the two dimensional AFM character in this compound[3,4]. This compound does not show any structural transition down to the lowest possible temperature 26K. Fig.6(b) shows the magnetic susceptibility measurement of x=0.2 which suggests that this compound goes paramagnetic to antiferromagnetic phase at a temperature ~ 47.5K followed by a broad peak at a temperature ~79K. The x=0.2 compound shows a more sharp peak at $T_N$ than x=0 compound which indicates that the magnetic 3D character is improved with the doping of Ru ions at Ti site in x=0 compound. Since we have doped the $Ru^{4+}$ magnetic ions at non-magnetic Ti site because of a clear

sharp anomaly has been observed at $T_N$. But in x=0 compound only Mn ion is magnetic. To understand the role of Ti ions in the magnetism of x=0, the Ru ion has been doped at Ti site. The magnetic susceptibility measurement results clearly show a sharp anomaly ~47.5K which comes after the Ru doping. The broad peak has been shifted towards the lower temperature and the value of χ has also been increased. This possibly may be because of the spin canting. A similar result has also been reported by Ramakrishnan *et al.* on the Nb doped $MnTiO_3$ at Ti site[9]. Goodenough et al. shows that x=0 compound does not show a sharp anomaly at $T_N$ due the accidental cancellation of the inter-layer exchange interactions[10]. But in x=0.2 compound we have doped the magnetic Ru ions at non-magnetic Ti site. The doping of this magnetic ion possibly will create more intra-and inter-layer exchange interactions. So here it may be that all the inter-layer exchange interactions are not getting cancelled by each other and net effect results in form of sharp anomaly at transition temperature. So it appears that the magnetic structure has been improved. The neutron diffraction study will be helpful to understand the magnetic structure of this system. It is interesting that the onset temperature for intra-layer antiferromagnetism is shifting towards the higher temperature even though the $T_{2D}$ is shifting towards the lower temperature as compared to parent compound. The shifting of intra-layer AFM onset temperature towards the higher temperature in x-ray diffraction study and an strong anomaly at ~ 47.5K in the magnetic susceptibility measurement is a signature of strengthening the intra-layer and inter-layer magnetic interactions in x=0.2 compound.

**Conclusion:**

In the x-ray diffraction study of x=0.2 compound it is observed that the lattice parameters a and c are linear up to 225K instead of 200K as in the case of x=0 compound. This is an indication of shifting the onset temperature for 2D antiferromagnetism towards the higher temperature. To ensure this fact we have subtracted the linear thermal contribution from the lattice parameters a and c. In the magnetic susceptibility measurement of x=0.2 compound the 2D and 3D magnetic transition temperatures have shifted towards the lower temperature by a temperature of 21K and 15.5K, respectively in comparison of the x=0 compound. In x=0 compound, the $Mn^{2+}$ ions are magnetic while $Ti^{4+}$ are non-magnetic. So the $Mn^{2+}$ ions are responsible for the magnetism in this compound. The role of $Ti^{4+}$ ions was unclear in this compound. To understand the role of $Ti^{4+}$ ions in x=0 compound the $Ru^{4+}$ magnetic ions have been doped at non-magnetic Ti site. The Ru doping induces the improved 3D and 2D magnetic character which results a sharp peak at a temperature ~47.5K and a broad peak at temperature ~79K in the dc magnetic susceptibility measurement for x=0.2 compound.

* *rkumar.0121@gmail.com*

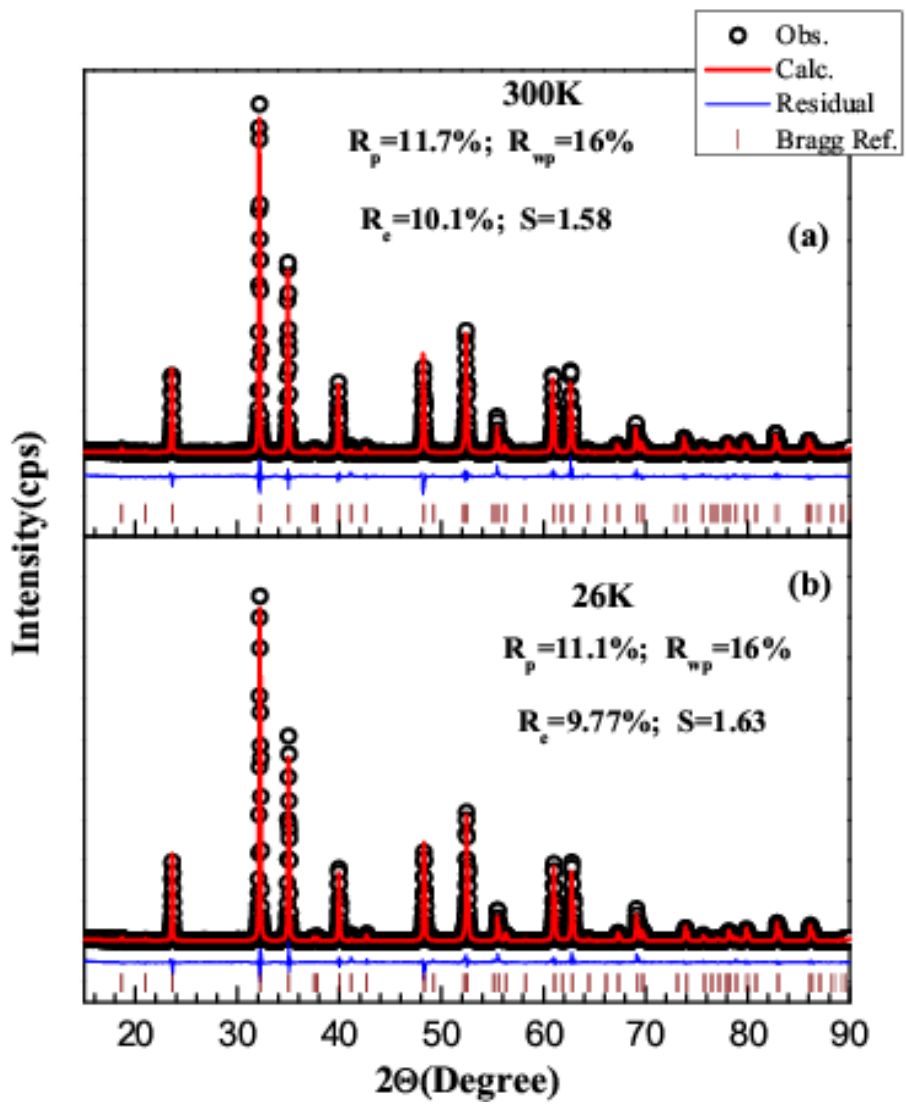

Fig.1: (colour online) Rietveld refinement of x=0.2 compound at (a) room temperature (300K) and (b) 26K. The open circles and the solid lines corresponds to the observed and calculated patterns, respectively. The vertical bars represent the position of Bragg reflection and the line shows the difference between observed and calculated intensities.

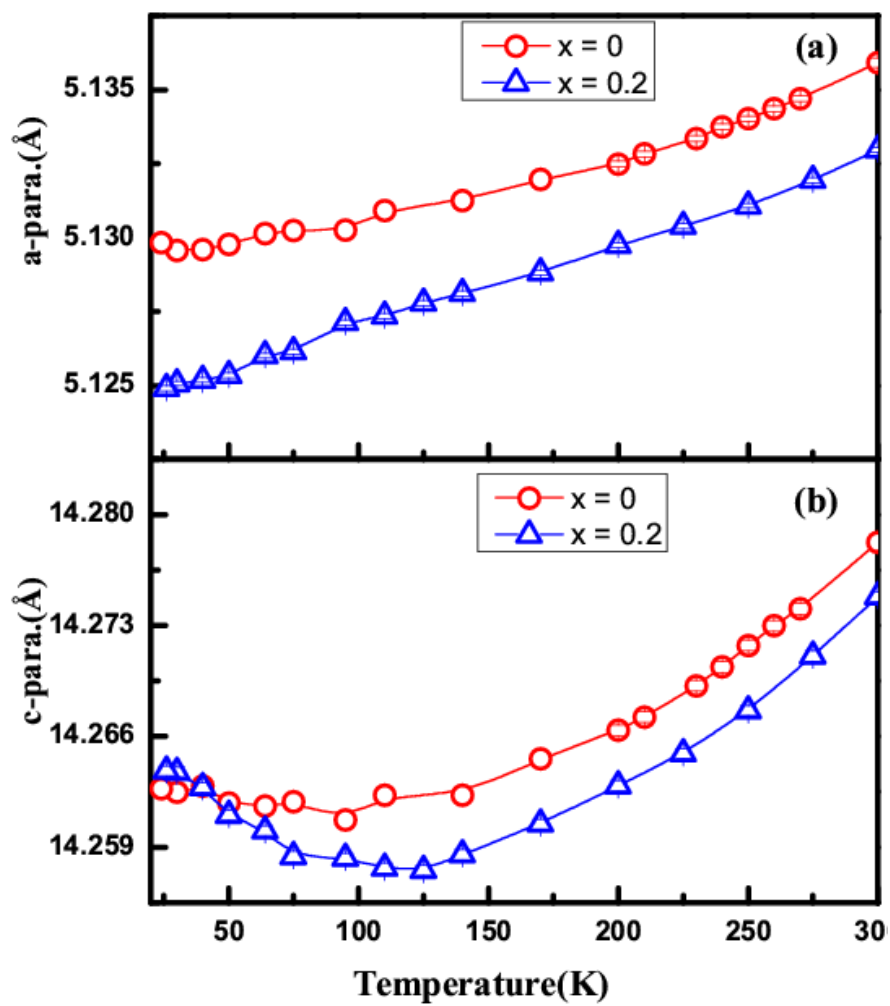

**Fig.2**: (Colour online) Temperature evolution of lattice parameters a and c for x=0 and x=0.2 compounds.

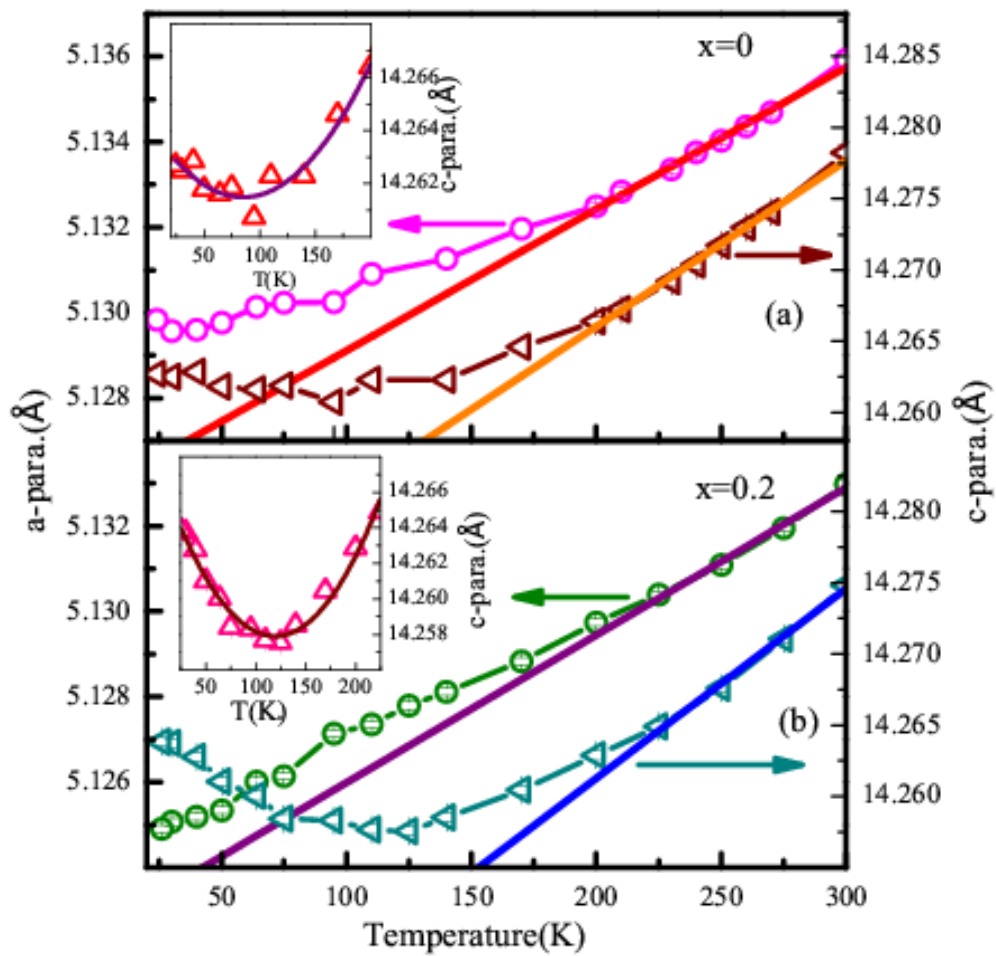

**Fig.3**:(Colour online) Temperature evolution of lattice parameters a and c for x=0 and x=0.2 compounds. The solid lines show the linear thermal contribution in the parameters. The inset figures show the closer view of c-parameters of x=0 and x=0.2 compounds.

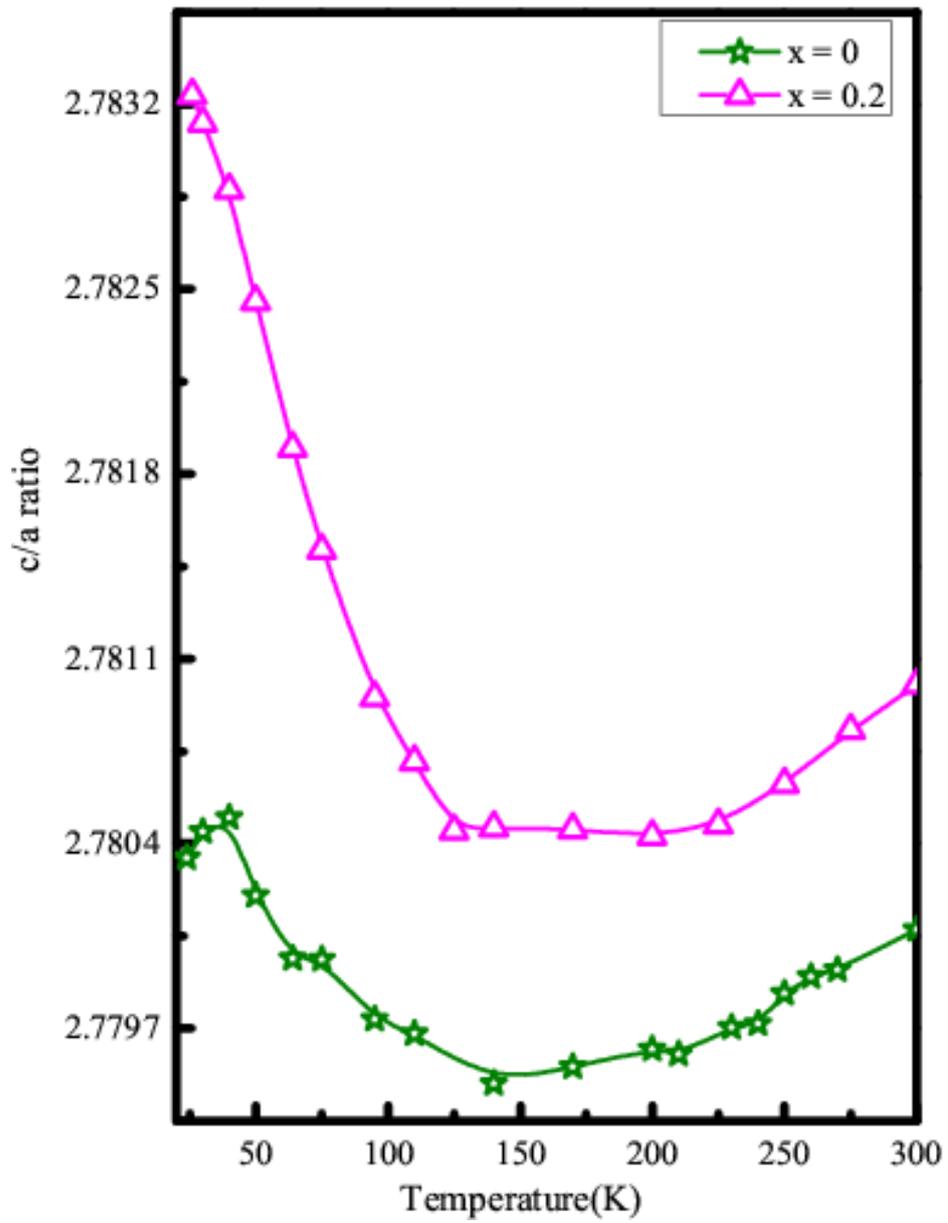

**Fig.4:** (Colour online )The temperature evolution of c/a ratio for x=0 and x=0.2 compound.

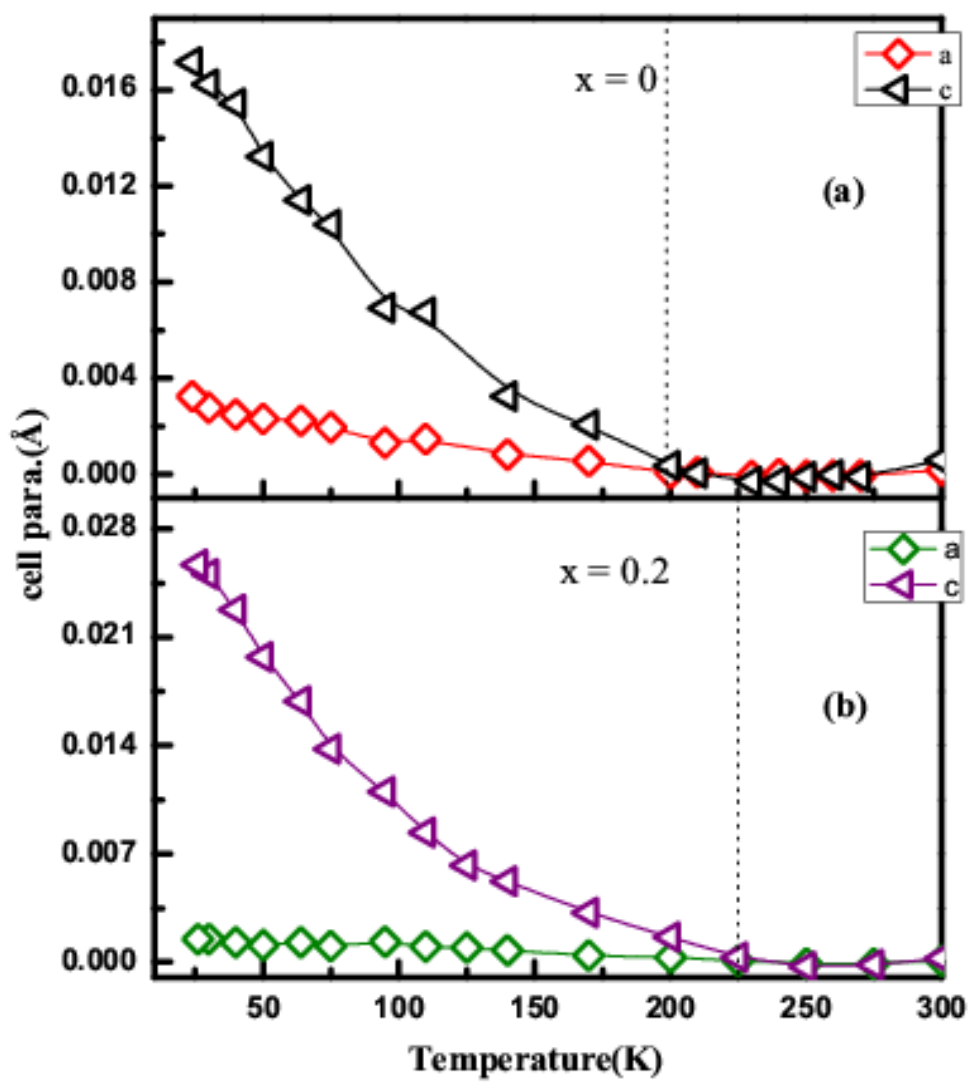

Fig.5: (Colour online)The temperature dependent behaviour of cell parameters after the subtraction of the linear thermal background of (a) x=0 and (b) x=0.2compound.

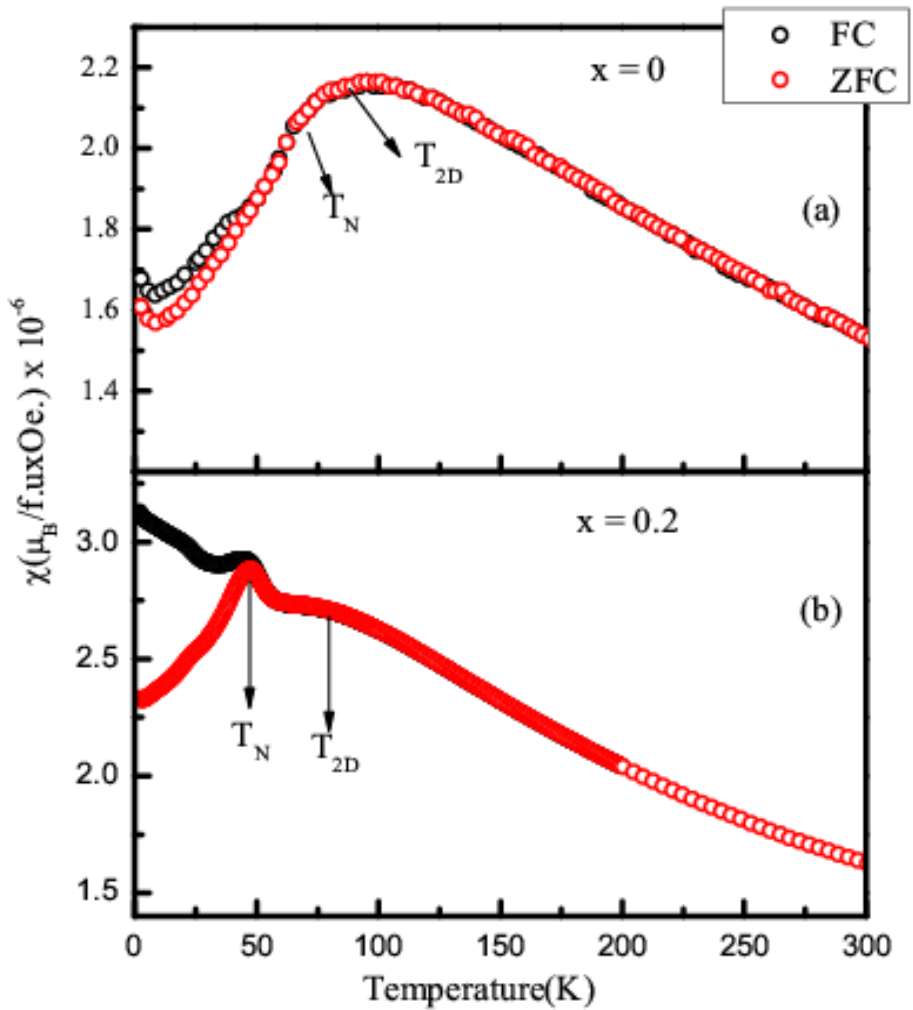

Fig.6: Zero field cooled (ZFC) and field cooled (FC) magnetic susceptibilities of x=0 and x=0.2 compounds at an external of 0.1 tesla. (a) $T_{2D}$ and $T_N$ for x=0 are ~ 100K and ~ 64K , respectively. (b) For x=0.2 $T_{2D}$ and $T_N$ are ~79K and ~47.5K , respectively.